\documentclass[a4paper]{article}

\usepackage{INTERSPEECH_v2}
\usepackage{amsmath}
\usepackage[nodisplayskipstretch]{setspace}
\usepackage{lscape}
\usepackage{dirtytalk,color}
\usepackage[dvipsnames]{xcolor}

\newcommand{\error}[1]{\textcolor{red}{\textbf{#1}}}

\title{Improving Performance of End-to-End ASR on Numeric Sequences}
\name{Cal Peyser$^{\star}$\thanks{$^\star$Contributed equally.}, Hao Zhang$^{\star}$, Tara N. Sainath, Zelin Wu}
\address{Google Inc., U.S.A}
\email{\{cpeyser,haozhang,tsainath,zelinwu\}@google.com}

\newcommand{\argmax}{\operatornamewithlimits{argmax}}

\AtBeginDocument{%
  \addtolength\abovedisplayskip{-0.55\baselineskip}%
  \addtolength\belowdisplayskip{-0.55\baselineskip}%
  \addtolength\abovedisplayshortskip{-0.55\baselineskip}%
  \addtolength\belowdisplayshortskip{-0.55\baselineskip}%
  \addtolength\abovecaptionskip{-0.55\baselineskip}%
  \addtolength\belowcaptionskip{-1.0\baselineskip}%
}

\begin{document}

\maketitle
\begin{abstract}
Recognizing written domain numeric utterances (e.g., I need \$1.25.) can be challenging for ASR systems, particularly when numeric sequences are not seen during training. This out-of-vocabulary (OOV) issue is addressed in conventional ASR systems by training part of the model on spoken domain utterances (e.g., I need one dollar and twenty five cents.), for which numeric sequences are composed of in-vocabulary numbers, and then using an FST verbalizer to denormalize the result. Unfortunately, conventional ASR models are not suitable for the low memory setting of on-device speech recognition. E2E models such as RNN-T are attractive for on-device ASR, as they fold the AM, PM and LM of a conventional model into one neural network. However, in the on-device setting the large memory footprint of an FST denormer makes spoken domain training more difficult.  In this paper, we investigate techniques to improve E2E model performance on numeric data. We find that using a text-to-speech system to generate additional numeric training data, as well as using a small-footprint neural network to perform spoken-to-written domain denorming, yields improvement in several numeric classes.  In the case of the longest numeric sequences, we see reduction of WER by up to a factor of 8.
\end{abstract}

\section{Introduction}
An ongoing challenge of ASR systems is to model transcriptions that do not exactly reflect the words spoken in an utterance.  For example, the spoken utterance \say{set an alarm for four fifteen} is typically decoded in the written form as \say{set an alarm for 4:15}.  Numeric utterances, such as addresses, phone numbers, and postal codes are particularly hard members of this category, due to the inherent out-of-vocabulary (OOV) issues of long written-domain numeric sequences. This problem arises because of a data sparsity issue, namely that long numeric sequences are unlikely to be present in training data.

A natural solution to the data sparsity/OOV problem is to train an ASR system in the spoken domain, and then to denorm the results back into the written domain \cite{shugrina,sak,chelba}.  Such a solution for conventional ASR systems was presented in \cite{Vasserman15}. In that work, the AM and PM are trained in the spoken domain. A verbalizer weighted finite state transducer (WFST) is then inserted before the LM to convert commonly used spoken-domain numerics to the written domain. Finally, a class-based n-gram LM, modeled as an FST, is used to transduce additional spoken domain entities to the written domain.

There are a few issues with the above approach for handling the numerics problem. First, it is based on a set of rules in the verbalizer and class-based LM, which does not scale well to changes in training data \cite{Zhang:EtAl:18}. In addition, the production FST-LM is large, which presents a challenge for memory-constrained on-device applications \cite{Ryan19}.

End-to-end (E2E) ASR models have gained popularity in recent years. These models, which fold the AM, PM and LMs into a single neural network, are particularly attractive for on-device applications. E2E models, including RNN-T, have been shown to achieve comparable performance to conventional models at a fraction of the size \cite{CC18,Ryan19}. However, these same constraints preclude the use of an FST denormer, and thus make spoken domain training more difficult.

The goal of this work is to improve E2E performance on numerics by applying several techniques. Synthetic data has been used successfully to address data sparsity issues in image processing for scene text recognition~\cite{jaderberg}, text spotting~\cite{gupta}, object detection~\cite{tremblay} and speech recognition~\cite{yanzhang}. Given this success, we first explore augmenting our training set using a text-to-speech (TTS) system to synthesize additional written-domain numeric training data.  In doing so, we choose transcripts from several challenging numerics categories in order to improve coverage of those cases.

Data-sparsity/OOV issues have also been addressed by neural correction models in both in NLP~\cite{bengio, sundermeyer} and in speech~\cite{kombrink, tilk,Jinxi19, tanaka, sproat}. A recent example in ASR is ~\cite{tanaka}, which proposed a neural error corrective language model (NECLM) trained on pairs of transcripts and corresponding errorful ASR output as a last step during decoding.  This approach was applied to denorming in ~\cite{sproat} with a neural written-to-spoken denormalizer. We build on the work in~\cite{sproat} by training a ~\say{neural correction} network, which is trained on written-domain ground-truth/RNN-T hypothesis pairs and learns to correct mistakes.

Given the success of the spoken domain for numeric entities, a third avenue we explore is to train RNN-T in the spoken domain and denorm back to written domain. We compare both a traditional FST-based denormer as in ~\cite{Vasserman15} and a ~\say{neural denormer} that is based on our written correction model. 

We evaluate the different E2E numeric solutions on a variety of different numeric test sets, which differ in their numeric sequence length. We find that while written domain approaches show significant improvements, the largest improvements are in the spoken domain, in which the OOV and data-sparsity issues are mitigated, and particularly for long numeric sequences.  Specifically, in the written domain, the addition of TTS training data gives a 65\% relative WER improvement for small numbers and a 35\% improvement for large numbers, increasing to 75\% and 49\% with the addition of a neural correction model. In the spoken domain, we see our best results, where the incorporation of a neural denorming model yields a 75\% improvement on short numbers and 83\% on large numbers.  On a curated eval set of specifically difficult verbalizations, results on long utterances are even stronger, showing a factor of 8 reduction in WER.

The rest of this paper is organized as follows.  Section \ref{sec:methods} introduces the four variations of standard RNN-T that we worked with.  This section includes our procedure for TTS utterance generation as well as our neural correction/denorm architecture.  Section \ref{sec:experiments} describes the details of our experiments.  Section \ref{sec:results} presents our results and analysis.  Finally, Section \ref{sec:conclusions} summarizes our conclusions.

\begin{center}
\begin{table*}[h!]
	\centering
	\begin{tabular}{|c|c|c|c|}
	\toprule
	Category     & Example Transcript Template         & Example Numeric & Average Numeric Length  \\
	\midrule
	DAY          & remind me on monday the \$DAY       & 31st  & 1.8           \\
	PERCENT      & turn down sound to \$PERCENT        & 20.22\% & 2.2         \\
	POSTALCODE   & how far away is \$POSTALCODE        & 86952  & 5.1      \\
	TIME         & set second alarm for \$TIME p.m.    & 10:46 & 3.0          \\
	YEAR         & play the top 40 from \$YEAR         & 1648  & 4.0        \\
	\bottomrule
	\end{tabular}
\vspace{0.2in}
\caption{Sample TTS Utterance Categories, with Example Synthetic Numerics}
\end{table*}

\end{center}

\section{Methods} \label{sec:methods}
In this section, we present different ideas explored to address the long-tail numeric issue of our RNN-T system.  We give each approach a label, which we will reference in Table 2 below.

\subsection{TTS Training Data (\texttt{W1})}
To address the numeric data-sparsity issue, we generate additional training data that represent challenging and realistic numeric sequences.
To this end, we choose numeric categories that we see often in Google Assistant traffic.  See Table 1 for sample categories.
We specifically choose categories that represent a variety of numeric sizes.
To obtain transcripts in these categories, we choose 200 unique templates for each category from anonymized Google Assistant utterances,
and inject the numeric entities by performing weighted sampling from our production numeric WFST grammar (weighted on the spoken domain).
Two parallel transcripts are generated for each template, one in the spoken domain and the other in the written domain. We perform this process 100 times for each template, each time synthesizing a unique utterance.
All models below use TTS training data generated in this manner.

\subsection{Written-Domain Neural Correction (\texttt{W2}) \label{sec:neuraldenorm}}
\begin{figure}[htb!]
  \includegraphics[width=1.2\columnwidth]{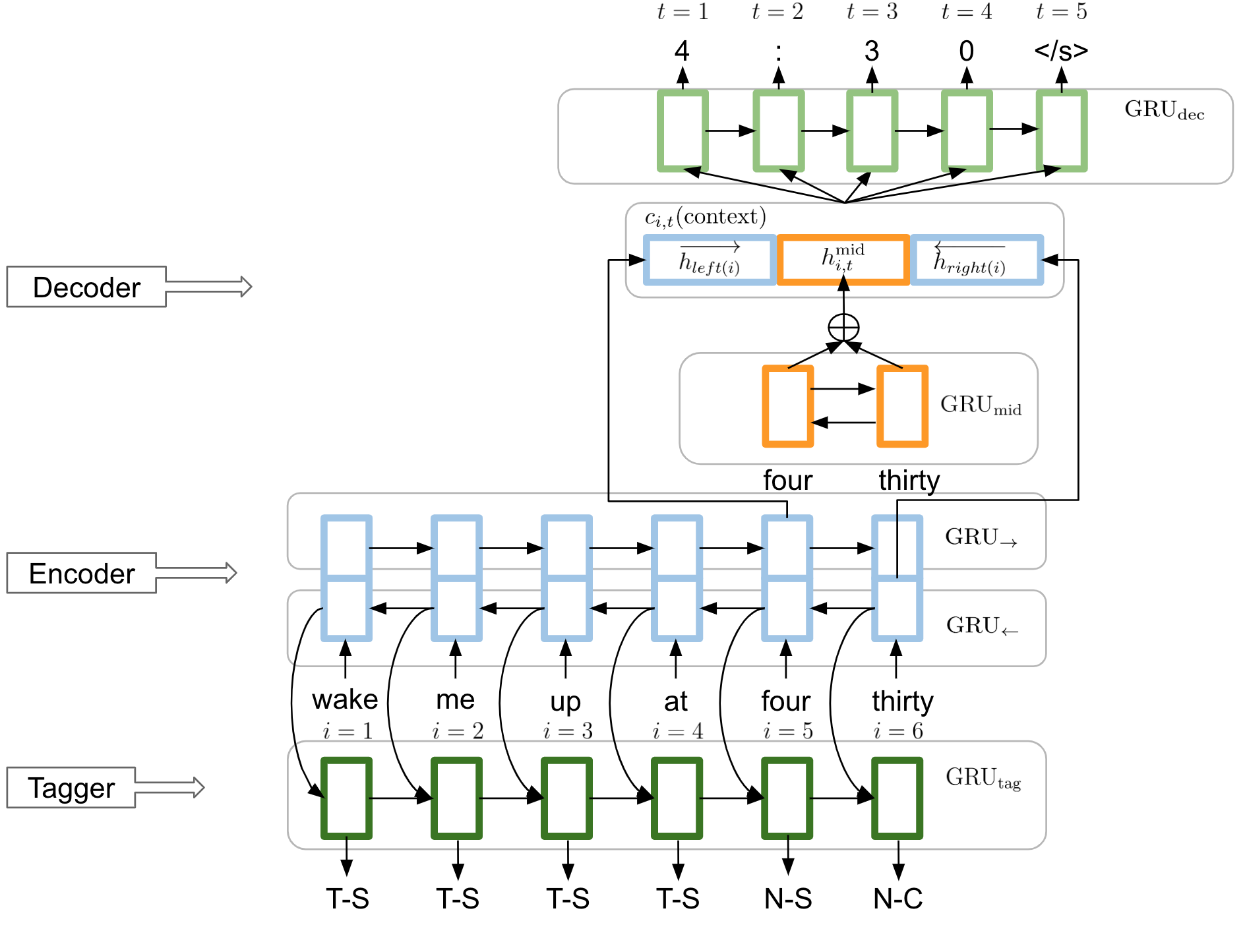}
  \caption{Neural Denormer Architecture. T stands for {\em trivial}. N stands for {\em non-trivial}. S stands for {\em start}. C stands for {\em continuation}.}
  \label{fig:denormer_arch}
\end{figure}

We seek to correct written RNN-T hypotheses with a post-processing neural correction step. Our written-domain correction model, shown in Figure \ref{fig:denormer_arch}, is an attention-based sequence-to-sequence model, for which the input is RNN-T hypotheses and the output is a corrected transcript.  Our architecture is adapted to the correction setting by accounting for the fact that the majority of words in an input sentence are simply copied to the output during correction.  This is done by training an additional ~\say{tagger} RNN to run on the input sequence before the sequence-to-sequence model.  The tagger tags each word in the input sequence as either ~\say{trivial}, in which case it is simply copied to the output sequence, or ~\say{non-trivial}, in which case it is passed into the attention model and decoder.  This architecture has been reported for contextual text normalization for text-to-speech ~\cite{Zhang:EtAl:18}, but we adapt it here to the correction setting for ASR.  While we use the model for correction, it could also be used to re-rank an n-best list in a second-pass setting.

\subsubsection{Encoder Layer}

Suppose we seek to map between input sequence $\mathbf{x} = \{x_1,\dots,x_I\}$ and output sequence $\mathbf{y} = \{y_1,\dots,y_T\}$, where the sequence vocabulary is composed of words. We define a BiRNN encoder such that

\[h_i = [\overrightarrow{h}_i;\overleftarrow{h}_i] \]
where
\[\overrightarrow{h}_i = \mathbf{RNN}_{\rightarrow}(\overrightarrow{h}_{i-1}, x_i) \]
and
\[\overleftarrow{h}_i = \mathbf{RNN}_{\leftarrow}(\overleftarrow{h}_{i+1}, x_i) \]
\\
where $\mathbf{h} = h_1,\dots,h_I$ are hidden encoder states.  

\subsubsection{Tagging Layer}

We define the \say{tagger} RNN as 

\[s_{i} = \mathbf{RNN}_{\text{tag}}(s_{i-1}, t_{i-1}, h_i) \]
\\
where $\mathbf{s} = s_i,\dots,s_I$ are hidden tagger states, with corresponding observations, i.e., tag sequence
$\mathbf{t} = t_i,\dots,t_I$.

Each tag $t_i$ is a joined tag in the cross-product set of $\{\mbox{trivial}, \mbox{non-trivial}\} \times \{\mbox{start}, \mbox{continuation}\}$ to model whether a word is the beginning of a new segment needed to be corrected or a continuation of the previous segment. This refinement allows us to model consecutive non-trivial segments.

The learning objective of the tagger is
\begin{equation}\label{eq:tag_objective}
 \argmax_{\mathbf{t}} \prod_{i=1}^{I}{P(t_i|s_i)}
\end{equation}
where $P$ is defined as a linear projection of $\mathbf{s}$ followed by a softmax layer.  We obtain alignments for training the tagger by using a heuristic alignment algorithm that determines subsequences common to the input and output transcripts.  These common subsequences are marked as ~\say{trivial}.

\subsubsection{Decoder Layer}

We use the results of the tagger to extract text snippets to be corrected.  Suppose a text snippet spans from time $s$  to $e$. The input span $\{x_{s},\dots,x_{e}\}$ along with the context hidden states $\overrightarrow{h}_{s}$ and $\overleftarrow{h}_{e}$ become the input to the next stage attentional model. We define a BiRNN encoder $\mathbf{RNN}_{\text{mid}}$ over $\{x_{s},\dots,x_{e}\}$.

Finally, the attentional RNN decoder $\mathbf{RNN}_{\text{dec}}$ is defined as
\[d_{i,t} = \mathbf{RNN}_{\text{dec}}(d_{i,t-1}, y_{i, t-1}, c_{i,t})  \]

where $c_{i,t}$ is the result of the attention function of $d_{i,t-1}$, $\overrightarrow{h}_{s}$, $\overleftarrow{h}_{e}$, and $\mathbf{RNN}_{\text{mid}}(\{x_{s},\dots,x_{e}\})$. The two-dimentional indices $(i,t)$ indicate $t$ is relative to a given position $i$ ($s$, $e$) in the input sentence.
The architecture of the attention component is the same as in ~\cite{Zhang:EtAl:18}.

The learning objective of the decoder is
\begin{equation}\label{eq:dec_objective}
 \argmax_{\mathbf{y}}{\prod_{t=1}^{L(i)}{P(y_{i,t}|d_{i,t})}}
\end{equation}




In this way, the attention mechanism and decoder are applied only to relevant spans of text, which decreases cost and improves accuracy. As shown in the derivation steps, at training time the two objectives in Equation~\ref{eq:tag_objective} and Equation~\ref{eq:dec_objective} translate to two cross-entropy losses that can be linearly combined during training. At decoding time, the two models work as a pipeline, with the attentional model only used as required by the tagging model.

\subsection{Spoken Domain Training, FST Denorming (\texttt{S1}) \label{sec:fst}}
Following the success in conventional ASR models, we explore spoken domain training for RNN-T. We train RNN-T on a spoken-domain version of our training set, and leave the translation back to written domain to an FST denormer derived from our production grammar \cite{Vasserman15}.

In order to train this model, we require examples of utterance transcripts in both the spoken and written domains.  We gather these examples by passing written-domain transcripts from our training set through an FST verbalizer.  We then choose a single verbalization by passing each candidate hypothesis through a lexicon, and force-aligning the resulting phone sequences against the phones in the utterance. For TTS training data, we simply use the spoken domain transcript that we obtained using our verbalization grammar.

\subsection{Neural Denorming (\texttt{S2})}
Since the FST-based denorming approach discussed in Section \ref{sec:fst} is challenging to put on device, we also explore a neural denorming approach. Specifically, we adapt our written correction model from Section \ref{sec:neuraldenorm} to the spoken domain by rephrasing it as a denormer which consumes spoken domain training data and emits written domain output. The architecture of the neural denorming model is identical to the written correction model.

\begin{table*}[t!]
	\centering
	\begin{tabular}
	{|c|c|c|c|c|c|c|}
	\toprule
	& & Written & Written TTS & Written TTS & Spoken TTS & Spoken TTS \\
	& &         &             & Neural Correction & FST Denorm & Neural Denorm \\
	\midrule
  & Average Num Len & \texttt{W0} & \texttt{W1} & \texttt{W2} & \texttt{S1} & \texttt{S2} \\
	\midrule
	VS & & 6.7 & 6.8 & 6.5 & 7.6 & 6.9 \\
	NUMERICS & 3.0 & 6.9 & 6.6 & 6.3 & 8.3 & 7.2 \\
	\midrule
	SAMPLED\_SHORT & 2.1 & 7.6 & 2.6 & 2.1 & 4.2 & 1.9 \\ 
	SAMPLED\_MEDIUM & 3.3 & 11.9 & 5.7 & 5.2 & 7.6 & 4.0 \\
	SAMPLED\_LONG & 7.4 & 26 & 16.9 & 13.2 & 16.8 & 4.4 \\
	\midrule
	TAIL\_SHORT & 2.7 & 6.8 & 4.9 & 4.7 & 7.3 & 4.2 \\
	TAIL\_MEDIUM & 4.3 & 15.4 & 12.9 & 11.6 & 11.6 & 6.7 \\
	TAIL\_LONG & 6.9 & 92.7 & 74.8 & 50.9 & 11.9 & 11.3\\
	\bottomrule
	\end{tabular}
\vspace{0.2in}
\caption{Results: WER by Model and Test Set}
\end{table*}

\section{Experiments}
\label{sec:experiments}

\subsection{Data Sets}

Our experiments are conducted on a $\sim$30,000 hour training
set consisting of 43 million English utterances. The training
utterances are anonymized and hand-transcribed, and are representative of Google’s voice search traffic in the United States.
Multi-style training (MTR) data are created by artificially corrupting the clean utterances using a room simulator, adding
varying degrees of noise and reverberation with an average SNR
of 12dB \cite{Chanwoo17}. The noise sources are drawn from YouTube and
daily life noisy environmental recordings. The real audio test sets
we report results on include $\sim$14.8K voice search (VS) utterances extracted from Google traffic, and real-audio numeric test set of $\sim$8.0K utterances derived from the VS distribution (NUMERICS).

In order to generate synthetic training data, we use a multi-speaker TTS system based on the baseline architecture as described in~\cite{weining},
where the Tacotron model~\cite{tacotron} generates a mel-spectrogram conditioned on phonemes and a 64-dimensional speaker embedding learned for each speaker during training.
The predicted mel-spectrogram is then inverted to time-domain waveform with a WaveRNN neural vocoder ~\cite{WaveRNN}. We again use MTR to add artificial noise to our synthesized audio.
The TTS training data consists of 84 English speakers covering American, Australian,
British and Singaporean accents with a Google Assistant clean speaking style, comprising 370 hours of audio in total.
During inference, numeric input texts are mapped to phonemes and a speaker is randomly selected.

This system is used to synthesize audio for templated transcripts, as described above.  In this manner we create training sets for each numeric category, and create our final training set by grouping all categories together.  For each batch in training, we draw 10\% of training examples from this TTS set and 90\% of examples from the real-audio training set described above.

Since we know that the length of a numeric sequence exacerbates the OOV issue, for our test sets we split numerics into three categories based on length.  For these categories, we generate transcripts using the same template phrases as in training, but with unique verbalizations, to ensure that no phrase is used in both training and test.  We call these sets the SAMPLED\_SHORT, SAMPLED\_MEDIUM and SAMPLED\_LONG test sets. We are also interested in evaluating our models on the ~\say{tail} of the distribution - that is, on the most difficult numerics.  To this end, for each category we maintain a list of particularly difficult numeric verbalizations.  For example, the postal code ~\say{22110} might be verbalized as ~\say{double two double one oh}.  We call the test sets generated with these more difficult verbalizations TAIL\_SHORT, TAIL\_MEDIUM, and TAIL\_LONG.

In order to ensure that our system has not overfit to a specific sort of TTS data, we synthesize audio for our test sets using separate TTS systems.  The audio for the SAMPLED sets is created using a single-speaker WaveNet ~\cite{WaveNet} system.  The audio for the TAIL sets is created using a concatenative TTS ~\cite{concat} system.

\subsection{RNN-T}
All experiments use the same RNN-T architecture, following the setup in \cite{Ryan19}. Specifically, an 8-layer unidirectional LSTM of width 2,048 with 640 recurrent projection units is used as the encoder.
A time-reduction layer with the reduction factor $2$ is inserted after the
second LSTM layer of encoder to reduce model latency. The RNN-T decoder consists of a prediction network with 2 LSTM layers of 2,048 hidden units and a 640-dimensional projection per layer as well as an embedding layer of 128 units. The outputs of the encoder and prediction network are fed to a joint network that
has 640 hidden units, followed by a 4,096 wordpiece softmax output.
In total, RNN-T has approximately 114M parameters.  The RNN-T model is implemented in
Tensorflow~\cite{AbadiAgarwalBarhamEtAl15} and trained on $8 \times 8$ Tensor Processing
Units (TPU) slices with a global batch size of 4,096 for $\sim 200K$ steps.



\subsection{Neural Correction/Denorming}
The neural denormer uses a bidirectional single-layer GRU encoder with 256 units that emits a 256-dimensional hidden state.  The tagging RNN is a single-layer GRU with 64 units.  The decoder is a bidirectional single-layer GRU with 256 units.  The encoder/tagger portion of the model, which runs for all input, contains about 2M parameters, while the attention/decoder portion of the model, which runs only for text spans marked for correction, contains about 4M parameters. The small footprint of the neural correction model makes it attractive for the on-device setting. The model is implemented in Tensorflow and trained asynchronously on 12 GPUs, with a batch size of 16.

\section{Results}
\label{sec:results}

Table 2 gives WER results for each of our experiments on the SAMPLED and TAIL test sets, as well as the real-audio VS and NUMERICS test sets.  We use the labels given in Section 2 (\texttt{W1}, \texttt{W2}, \texttt{S1}, \texttt{S2}) for convenience. We use \texttt{W0} to refer to the baseline RNN-T model.

The results for the written domain models are characterized by a steep decline in quality with increasing numeric length.  While the the addition of TTS data in \texttt{W1} and the incorporation of a correction model in \texttt{W2} improve over the baseline \texttt{W0}, the error rates are still sometimes higher on the MEDIUM and LONG test sets than on the non-numeric VS set. 

The errors made by model tell a clear story: the written domain models struggle to correctly format numerics, and often confuse parts of numeric sequences for similar-sounding words. The following are representative errors taken from the SAMPLED sets.  Here, references are on the left of the arrow and erroneous hypothesis are on the right.
\\
\texttt{W1}: \begin{center} \say{\$180.50 into inr} $\rightarrow$ \say{\error{\$180 - \$50} in inr} \end{center}
\begin{center} \say{house for rent 60003} $\rightarrow$ \say{house for rent \error{\$6003}} \end{center}
\texttt{W2}: \begin{center} \say{\$487 / 6} $\rightarrow$ \say{\error{4876}} \end{center}
\begin{center} \say{48007 to 08618} $\rightarrow$ \say{\error{480-708-6618}} \end{center}

In the spoken domain, we see a different error pattern.  \texttt{S1} errors show that the FST denormer often completely or partially fails to perform denorming, for example: 
\\
\texttt{S1}: \begin{center} \say{code 30441} $\rightarrow$ \say{code \error{300 double 41}} \end{center}
\begin{center} \say{the 32nd door} $\rightarrow$ \say{the \error{30 second door}} \end{center}

This indicates the problem with rule-based FST denormers, which often cannot adapt well to changes in training data.

Progressing to \texttt{S2}, these issues largely disappear, and the errors that remain are a blend of formatting, denorming, and other errors, in considerably smaller numbers than for all other models.  It seems that the avoidance of OOV issues obtained by training in the spoken domain largely solves the formatting problems experienced by the written domain models, while using a neural denormer, which learns how to denorm from data, sidesteps the denorming errors seen in the FST-based spoken domain model. Finally, the spoken denorming approach does not result in a significant degradation on the real-audio VS or NUMERIC sets.

\section{Conclusions}
\label{sec:conclusions}

In this paper, we experimented with four approaches for improving end-to-end ASR performance on numeric utterances.  We found that all approaches yield improvements, with the largest improvements occurring when TTS training data, spoken domain training, and neural denorming are all used together.  The fact that we see the largest improvements in the longest utterances, together with the types of errors that seem to be avoided, suggest that the improvements are mainly due to the alleviation of the out of vocabulary problem.

\section{Acknowledgements}
We thank Gabriel Mechali, Mark Epstein, Michael Riley, and Richard Sproat for help and comments on this work.
\label{sec:ack}

\bibliographystyle{IEEEtran}
\bibliography{main}
\end{document}